\documentclass[aps,prd,twocolumn,showpacs,superscriptaddress,groupedaddress,nofootinbib]{revtex4}  
\usepackage{graphicx}  
\usepackage{dcolumn}   

\usepackage{graphicx}
\usepackage{amsmath}
\usepackage{amsfonts}
\usepackage{epsfig}
\usepackage{slashed}
\usepackage{braket}

\usepackage{comment}
\usepackage{cases}

\def\bec{\begin{center}}
\def\eec{\end{center}}
\def\beq{\begin{equation}}
\def\eeq{\end{equation}}
\def\bea{\begin{eqnarray}}
\def\eea{\end{eqnarray}}

\begin{document}
\title{Symmetric Mass Generation in Lattice Gauge Theory}
\author{Nouman Butt}
\affiliation{Department of Physics, University of Illinois at Urbana-Champaign, 1110 W Green St, Urbana, IL 61801} 
\author{Simon Catterall}
\affiliation{Department of Physics, Syracuse University, Syracuse, NY 13244, USA }
\author{Goksu Can Toga}
\affiliation{Department of Physics, Syracuse University, Syracuse, NY 13244, USA }

\date{\today}
\begin{abstract} We construct a four dimensional lattice gauge theory in which fermions acquire mass without breaking symmetries as a result of gauge interactions. Our model consists of
reduced staggered fermions transforming in the bifundamental representation of a $SU(2)\times SU(2)$ gauge symmetry. This fermion representation ensures that
single site bilinear mass terms vanish identically. A symmetric 
four fermion operator is however allowed and we
show numerical results that show that 
a condensate of this operator develops in the vacuum. 
\end{abstract}

\maketitle

\section{Introduction}
Can we generate a mass for all physical states in a
theory without breaking symmetries ? And can we do it using just gauge
interactions? In this paper we describe a lattice model which is capable of realizing this scenario.

The model consists of a reduced staggered fermion coupled to an
$SU(2)\times SU(2)$ lattice gauge field. The fermion representation is chosen
so that single site fermion gauge invariant fermion bilinear terms
vanish identically. A symmetric four fermion operator remains invariant under
these symmetries and we present
evidence that it condenses as a result of the gauge interactions. 
Since no symmetries are broken by this 
condensate there are no massless Goldstone bosons and the
spectrum consists of color singlet composites of the elementary fermions.
This scenario corresponds to symmetric mass generation realized
in the context of a confining gauge
theory. It gives an explicit and non-supersymmetric
realization of a mechanism that has been proposed to gap chiral
fermions in the continuum  \cite{Razamat:2020kyf,Tong:2021phe}. 

The model can be seen as a generalization of the $SO(4)$ Higgs-Yukawa theory described in \cite{Butt:2018nkn} which uses strong quartic interactions to gap lattice fermions. This four dimensional model built on earlier work directed at symmetric mass generation
with staggered fermions in two, three and four dimensions \cite{Ayyar:2014eua,Catterall:2015zua,Ayyar:2015lrd,Ayyar:2016lxq,Ayyar:2017qii,
Catterall:2020fep,Butt:2021brl}. 

In the current paper both of
the $SU(2)$ subgroups of $SO(4)$ are gauged and confinement rather than
strong Yukawa interactions is used to generate the four fermion 
condensate. 

\section{Staggered Fermion Model}
We start from a staggered fermion action which takes the form
\begin{equation}
    S=\sum_{x,\mu}\eta_\mu(x){\rm Tr}\,\left(\psi^\dagger \Delta_\mu \psi\right)
    \label{free}
\end{equation}
where $\eta_\mu(x)=\left(-1\right)^{\sum_i^{\mu-1} x_i}$ are the usual
staggered phases and $\Delta_\mu$ is the symmetric difference operator whose action on a
lattice field $f(x)$ is given by
\begin{equation}
    \Delta_\mu f(x)=\frac{1}{2}\left(f(x+\mu)-f(x-\mu)\right)
\end{equation}
This action has a $U(1)$ staggered symmetry $\psi\to e^{i\epsilon(x)\alpha}\psi$ with $\epsilon(x)=\left(-1\right)^{\sum_i x_i}$ the site parity operator.
The fermions are additionally taken to transform under a {\rm global} $G\times H$ symmetry where $G$ and $H$ correspond to
independent $SU(2)$ groups:
\begin{equation}
\psi\to \hat{\psi}=G\psi H^\dagger
\end{equation}
We will also impose the reality condition 
\begin{equation}
    \psi^\dagger=\sigma_2\psi^T\sigma_2
\end{equation}
To see that this reality condition is compatible with the
non-abelian symmetry consider the transformed fermion:
\begin{align}
    \hat{\psi}^\dagger&=H\psi^\dagger G^\dagger\\\nonumber
    &=H\sigma_2\psi^T\sigma_2 G^\dagger\\\nonumber
    &=\sigma_2\left(\sigma_2 H\sigma_2 \psi^T \sigma_2 G^\dagger\sigma_2\right)\sigma_2\\\nonumber
    &=\sigma_2 \left(H^*\psi^T G^T\right)\sigma_2\\\nonumber
    &=\sigma_2 \hat{\psi}^T\sigma_2
\end{align}
The reality condition is automatically satisfied if $\psi=\sum_{A=1}^4 \sigma_A \chi_A$
for real $\chi_A$ where $\sigma_A=\left(I,i\sigma_i\right)$. Substituting this expression
into the kinetic term shows that the action can be written in an 
explicit $SO(4)$ invariant form
\begin{equation}
    S=\sum_{x,\mu} \frac{1}{2}\eta_\mu(x)\chi^A \Delta_\mu \chi^A
\end{equation}
Indeed in this form one can see that the kinetic term of this model is precisely the
same as that considered in previous work with $SO(4)$ invariant staggered fermions \cite{Butt:2018nkn}.

Once this reality condition is imposed it is not possible to write down single
site mass terms since
${\rm Tr}\,\left(\psi^\dagger\psi\right)={\rm Tr}\,\left(\sigma_2\psi^T\sigma_2\psi\right)=0$ on account of
the Grassmann nature of the fields.
However a four fermion term 
invariant under $SO(4)=SU(2)\times SU(2)$ is possible and takes the form
\begin{equation}
{\rm Tr}\,\left(\psi\psi^\dagger\psi\psi^\dagger\right)=\frac{1}{3}\epsilon_{abcd}\chi^a\chi^b\chi^c\chi^d
\label{fourfermi}
\end{equation}
The form of this four fermion term also agrees with the earlier work \cite{Butt:2018nkn}.

To add such a four fermion term to the action
we can use a Yukawa interaction with a auxiliary scalar field
$\phi$ of the form
\begin{equation}
    \sum_x {\rm Tr}\,\left(\phi\psi\psi^\dagger\right)+\frac{1}{2\lambda^2}\sum_x {\rm Tr}\,(\phi^2)
    \label{yukawa}
\end{equation}
where $\phi$ transforms in the adjoint representation of $G$ but is a singlet under $H$:
\begin{equation}
 \phi\to G\phi G^\dagger   
\end{equation}
After integration over $\phi$ a four fermion term is produced with coupling $-\lambda^2/2$.
The addition of this term breaks the original $U(1)$ staggered symmetry to a $Z_4$ corresponding to
\begin{equation}
    \psi(x)\to \omega\epsilon(x)\psi(x)
\end{equation}
where $\omega$ is an element of $Z_4$

In \cite{Butt:2018nkn} we showed that it was possible to achieve symmetric mass generation in this
model for large values of the Yukawa coupling and vanishing gauge coupling.
In this paper we will show that a four fermion
condensate 
can also be obtained by using strong gauge interactions and small Yukawa coupling.
This result is important as it avoids the problem of relying on perturbatively irrelevant 
four fermion operators to induce symmetric mass generation.

To do this we need to generalize eqn.~\ref{free} so that it is invariant under
lattice gauge transformations. The following prescription does the job:
\bea
\label{covdif}
S_F = \sum_{x,\mu} & \frac{1}{2} \eta_{\mu}(x){\rm Tr}\, [ \psi^{\dagger}(x) U_{\mu}(x) \psi(x+\mu)V_{\mu}^{\dagger}(x) \\ \nonumber
&- \psi^{\dagger}(x) U^{\dagger}_{\mu}(x-\mu)\psi(x-\mu) V_{\mu}(x-\mu)]
\eea
 In addition we will add Wilson terms for $G$ and $H$:
 \begin{align}
     S_W=&-\frac{\beta_G}{2}\sum_{x}\sum_{\mu\nu}\left(U_\mu(x)U_\nu(x+\mu)U_\mu(x+\nu)^\dagger U_\nu(x)^\dagger\right)\\
         & -\frac{\beta_H}{2}\sum_{x}\sum_{\mu\nu}\left(V_\mu(x)V_\nu(x+\mu)V_\mu(x+\nu)^\dagger
     V_\nu(x)^\dagger\right)
 \end{align}
The resultant action is now invariant under the following gauge transformations
\label{eq:x}
\begin{align}
\psi(x) &\to G(x) \psi(x) H^{\dagger}(x) \\
U_{\mu}(x) &\to G(x)U_{\mu}(x) G^{\dagger}(x+\mu) \\ 
V_{\mu}(x) &\to H(x+\mu) V_{\mu}(x) H^{\dagger}(x) 
\end{align}
The Yukawa interaction given in eqn.~\ref{yukawa} is automatically invariant under
these local symmetries.~\footnote{As an aside we remark
that four fermion interactions similar to the ones considered here have previously been used to argue for the appearance of Higgs phases in strongly coupled lattice theories \cite{Catterall:2013sto,Catterall:2014zha}. }

Finally we note that the action of the model is
invariant under a $Z_2$ center symmetry transformation
\begin{align}
    V_\mu(x)&\to -V_\mu(x)\\\nonumber
    U_\mu(x)&\to U_\mu(x)\\\nonumber
    \psi(x)&\to \epsilon(x)\psi(x)
\end{align}
The existence of an exact center symmetry ensures that the Polyakov line 
\begin{equation}P(x)=\frac{1}{2}{\rm Tr}\,\prod_{t=1}^L V_\mu(x+t)\end{equation} is a good order parameter for confinement in this theory.

In the next section  we will show numerical results that provide evidence that a four
fermion condensate appears in the theory even for small
Yukawa coupling. 
We can think of this condensate as corresponding to the 
appearance of a bilinear mass term for the color singlet composite scalar $\phi=\psi\psi^\dagger$.
This scenario is similar to that advocated for
by Tong et al. in \cite{Razamat:2020kyf} as a mechanism for gapping chiral fermions. It is important to remember
though that this model targets a vector-like theory at short distances
as $\beta\to\infty$.

\section{Numerical results}

The fermion kinetic term including the Yukawa term takes the form
\begin{equation}
    S_F=\sum_x{\rm Tr}\,\left[\sigma_2\psi^T\sigma_2\left(
    \eta_\mu(x)\Delta^c_\mu+G\phi\right)\psi\right]
\end{equation}
where $\Delta^c$ denotes the covariant difference operator appearing in eqn.~\ref{covdif}.
Using the properties
\begin{align}
    U^T_\mu&=\sigma_2 U^\dagger_\mu \sigma_2\\\nonumber
    V^T_\mu&=\sigma_2 V^\dagger_\mu \sigma_2\\\nonumber
    \phi^T&=-\sigma_2\phi\sigma_2
\end{align}
It is easy to verify that the fermion 
operator $M$ is antisymmetric and each eigenvector $v_n(x)$ with eigenvalue $\lambda_n$
is paired with another $\sigma_2v_n^*(x)\sigma_2$ with eigenvalue $\lambda^*_n$.
Thus the eigenvalues, which are generically complex, come in
quartets $ \left(\lambda , \bar{\lambda} , -\lambda ,- \bar{\lambda}\right) $. This
ensures that the Pfaffian that results after fermion integration is generically real
positive definite. We have verified that this is indeed the case by computing the latter for
ensembles of small lattice size~\footnote{Notice though that 
pure imaginary eigenvalues
come only in pairs which allows for a sign change if such an eigenvalue crosses
the origin. While this is logically possible we have not seen any sign of
this in our simulations.}. Thus the model
can be simulated using the RHMC algorithm \cite{Catterall:2018pms,Butt:2018yxr}.
We now turn to our numerical results.

\subsection{The Yukawa theory limit $\beta_H=\beta_G \to \infty$}
\begin{figure}[htb]\centering
    \includegraphics[width=0.45\textwidth]{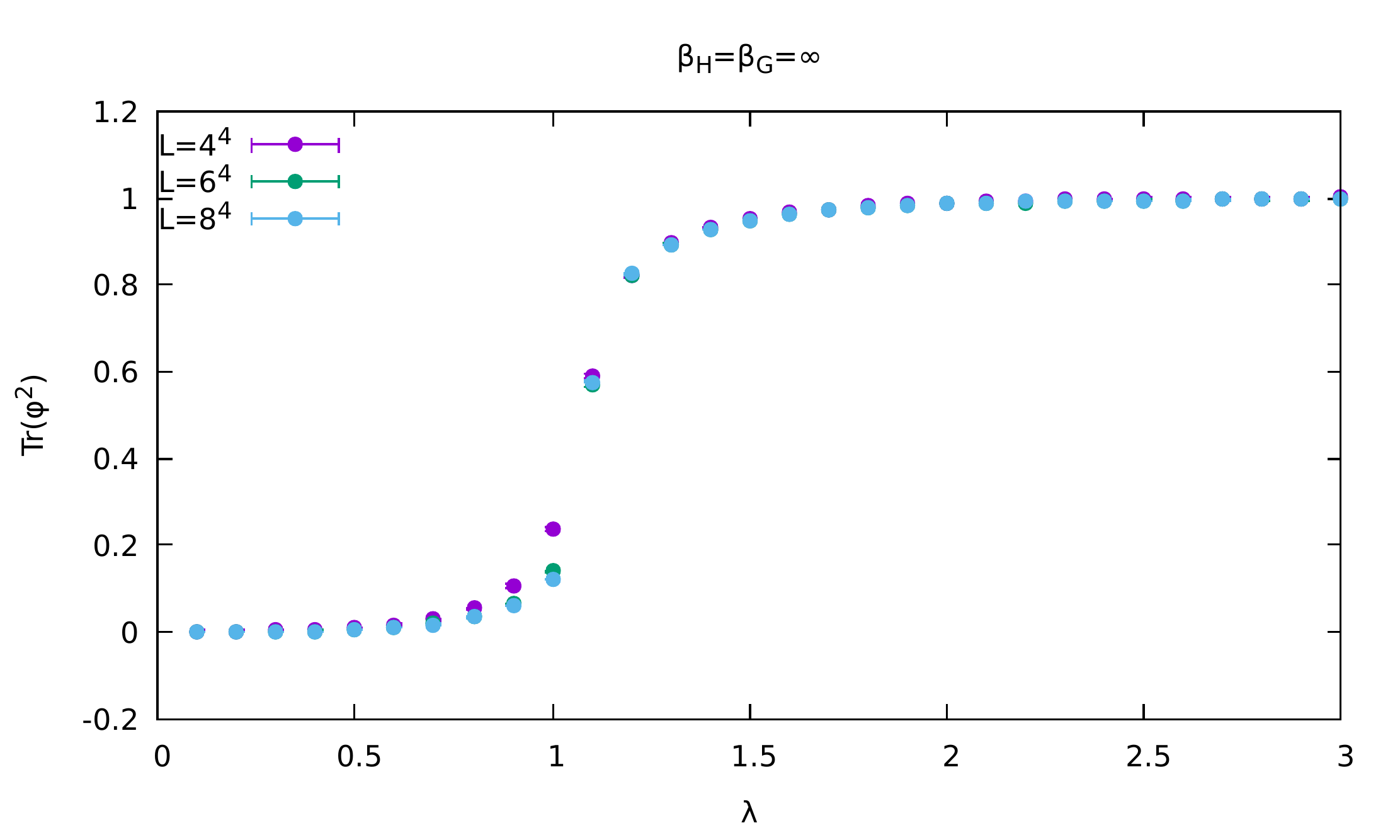}
    \caption{${\rm Tr}\,\left(\phi^2\right)$ vs $\lambda$ for  $L=4^4,6^4,8^4$}
    \label{sigmasq_not_gauged}
\end{figure}

\begin{figure}[htb]\centering
    \includegraphics[width=0.45\textwidth]{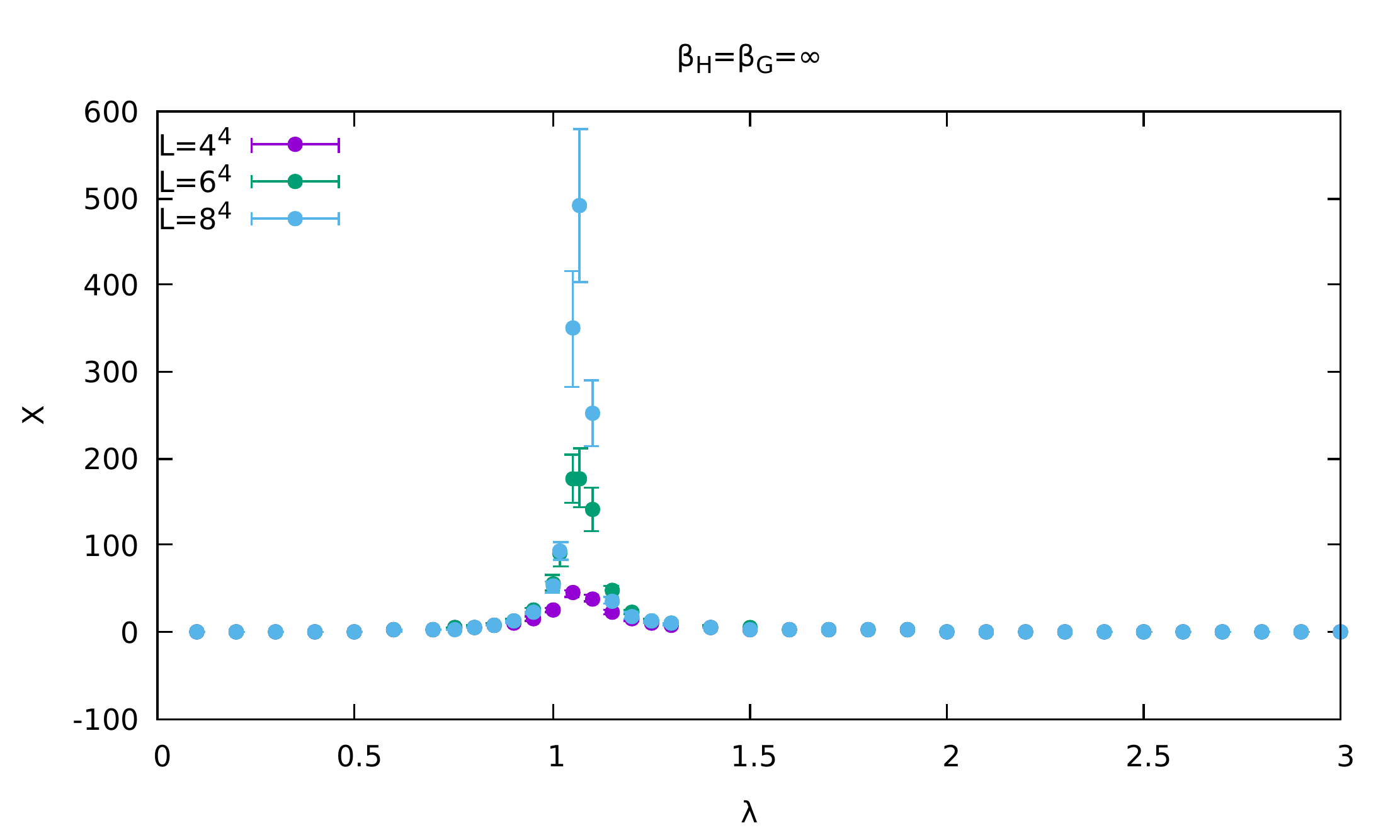}
    \caption{$<\chi>$ vs $\lambda$ for  $L=4^4,6^4,8^4$}
    \label{sus_not_gauged}
\end{figure}

In the absence of gauge interactions 
the model reduces to the $SO(4)$ Higgs-Yukawa theory
examined in \cite{Butt:2018nkn}.
In this limit the only way to drive a four fermion condensate
is through the use
of a large Yukawa coupling $\lambda$. Fig~\ref{sigmasq_not_gauged} shows a picture of the 
${\rm Tr}\,\phi^2$ which serves as a proxy for the four
fermion condensate vs $\lambda$. The rapid growth near $\lambda \sim 1.0$
is identical to our earlier results for the pure four fermion model in four
dimensions. This conclusion is strengthened in fig.~\ref{sus_not_gauged} which shows a plot of the
associated fermion
susceptibility defined by
\begin{equation}
\chi=\frac{1}{V}\sum_x\left\langle \psi^\dagger(0)\psi(0)\psi^\dagger(x)\psi(x)\right\rangle
\end{equation}
This shows a peak that grows with volume close to the coupling where the enhanced four
fermion vev switches on.

\subsection{The gauge theory limit $\lambda\to 0$}
\begin{figure}[htb]\centering
    \includegraphics[width=0.45\textwidth]{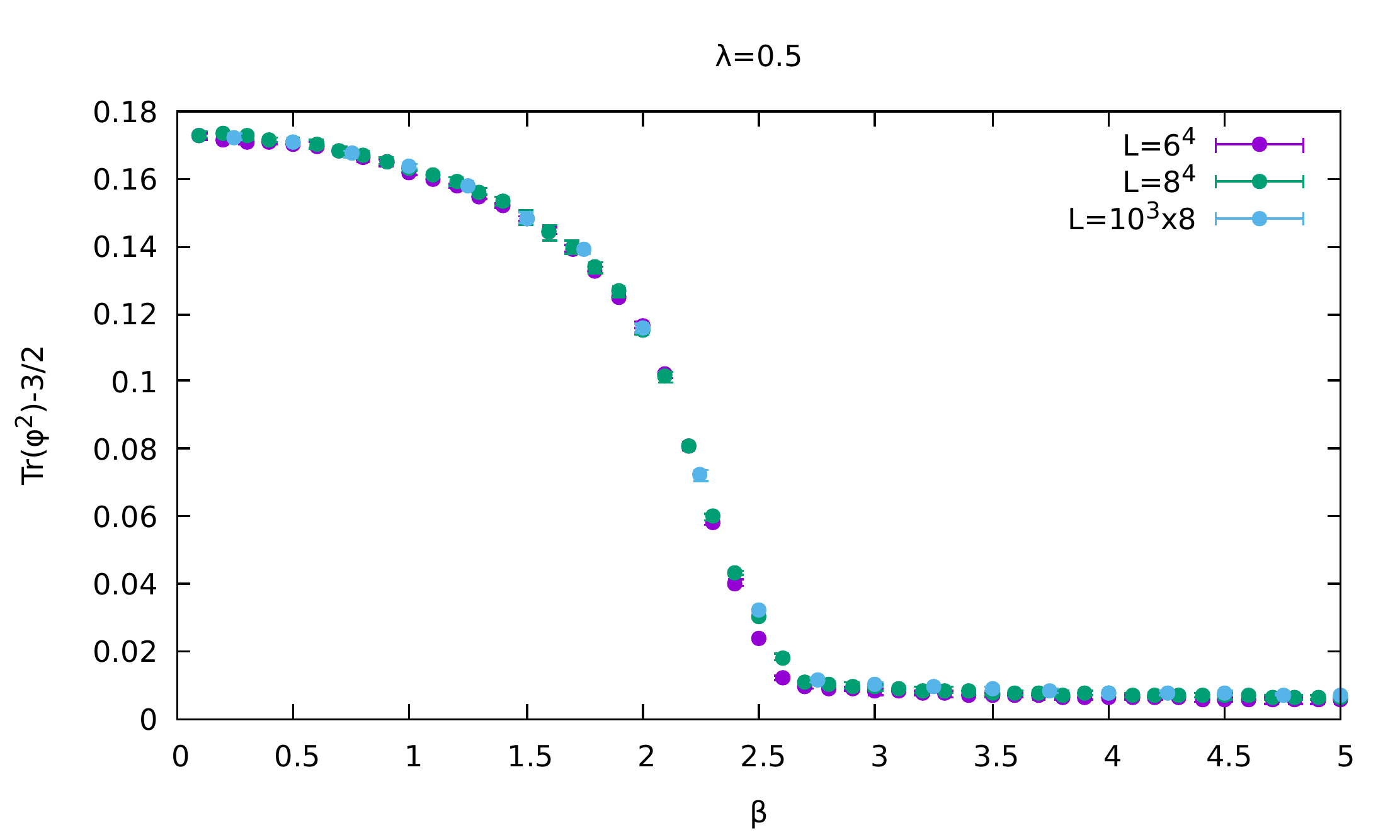}
    \caption{${\rm Tr}\,\left(\phi^2 \right)$ vs $\beta$ for  $L=6^4,8^4,10^3\times 8$}
    \label{betascan}
\end{figure}
\begin{figure}[htb]\centering
    \includegraphics[width=0.45\textwidth]{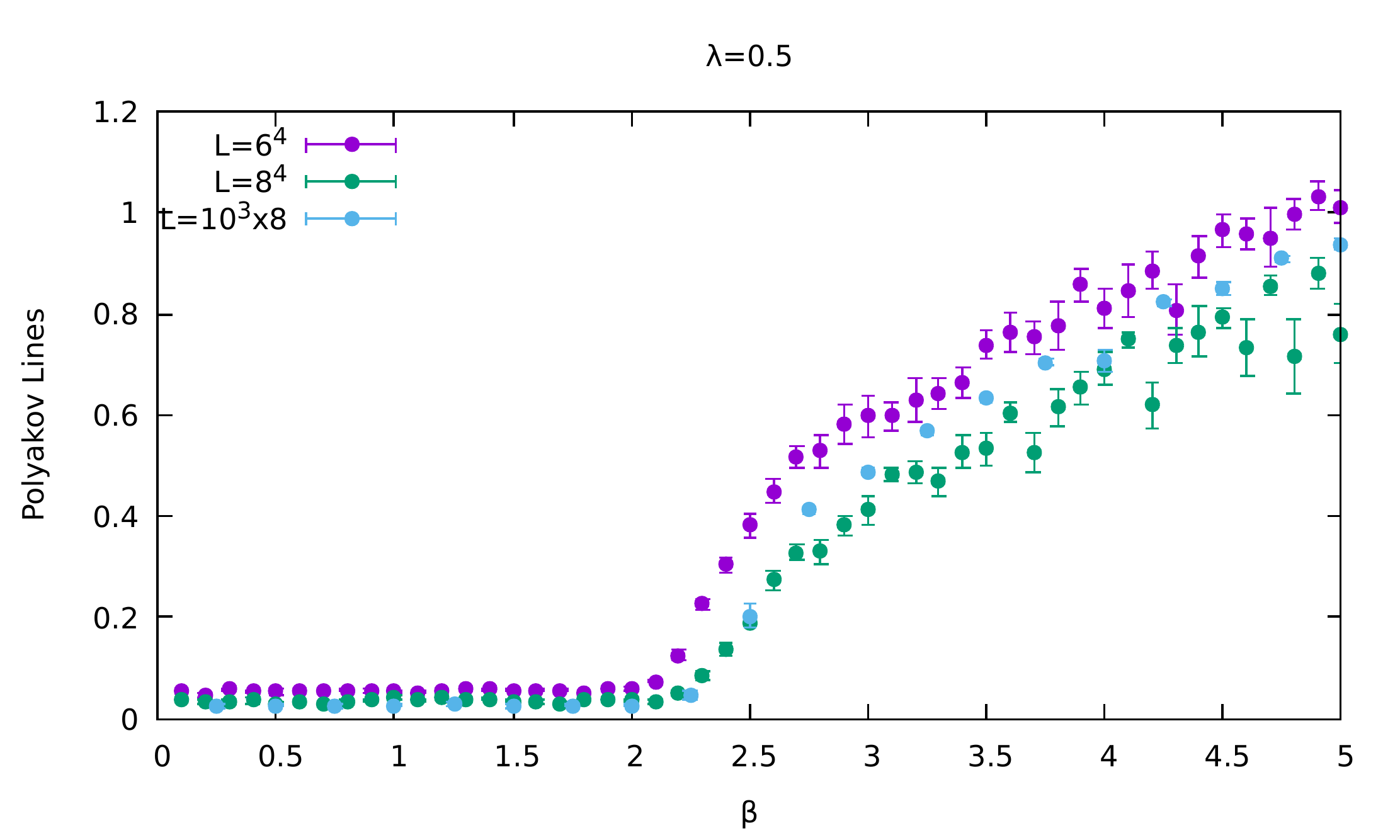}
    \caption{Polyakov line vs $\beta$ for  $L=6^4,8^4, 10^3\times 8$}
    \label{poly}
\end{figure}

\begin{figure}[htb]\centering
    \includegraphics[width=0.45\textwidth]{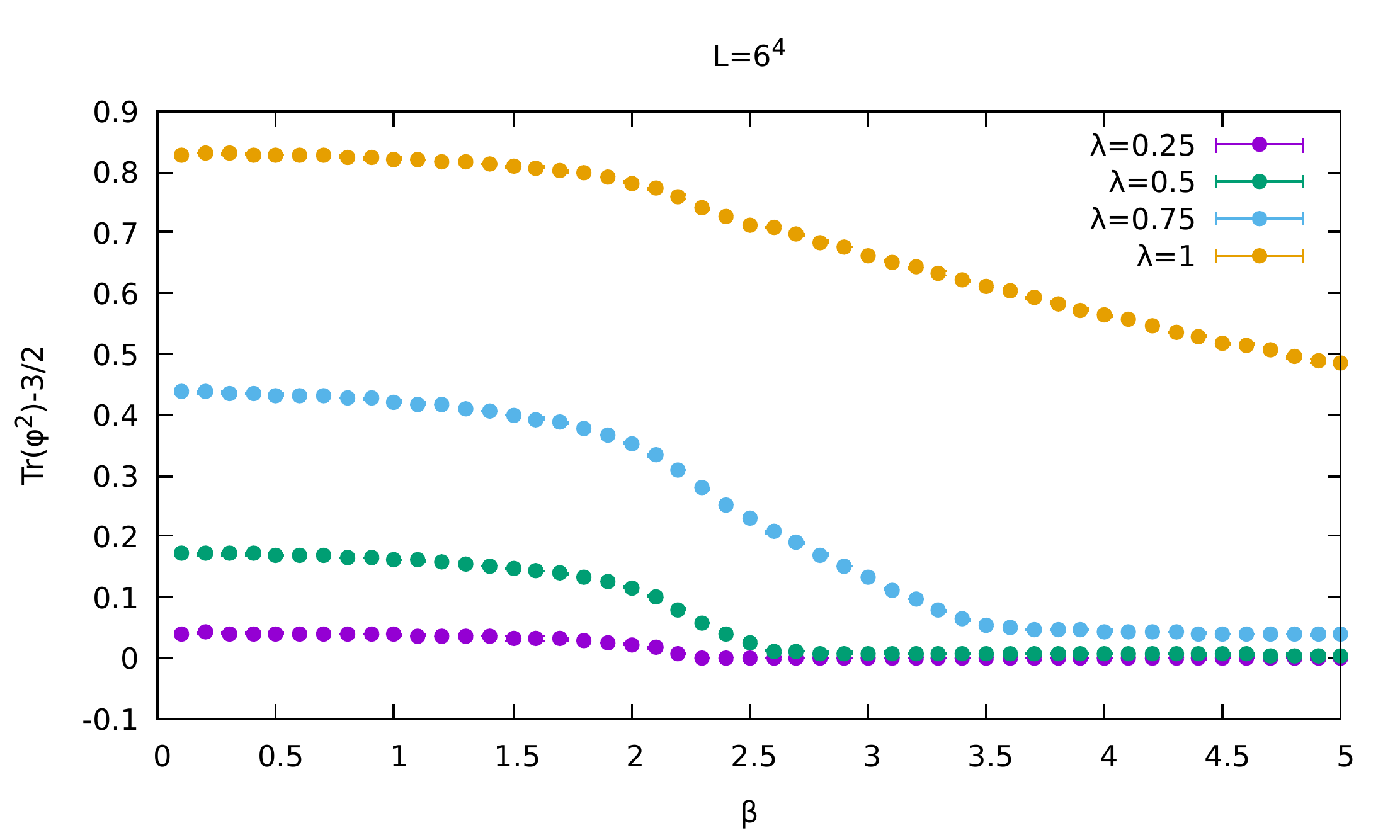}
    \caption{${\rm Tr}\,\left(\phi^2\right)$ vs $\beta$ for  $\lambda=0.25,0.5,0.75,1.0$}
    \label{sigmasq_L6}
\end{figure}

We now switch on the gauge interactions, setting  $\beta_H=\beta_G=\beta$ and retaining only a small Yukawa coupling.
Fig.~\ref{betascan} shows a plot of $Tr(\phi^2)$ vs $\beta$. The Yukawa coupling is small and fixed to $\lambda=0.5$ for lattice sizes $L=6^4,8^4, 10^3\times8$. Clearly the condensate grows for $\beta\le 2.5$. Notice that the appearance of this
four fermion vev is a result of the gauge interactions not the
explicit Yukawa coupling since the latter lies well below the threshold to drive the phase
transition seen in fig.\ref{sigmasq_not_gauged}. Indeed the effect of
changing the value of the bare Yukawa coupling $\lambda$ can be
seen in fig.\ref{sigmasq_L6} which shows the condensate
for a range of $\lambda=0.25,0.5,0.75,1.0$  on an $L=6^4$ lattice. Clearly
for all $\lambda<0.8$ a condensate develops at small $\beta$ but is
driven to zero in the weak gauge coupling limit $\beta\to\infty$
consistent with fig.\ref{sigmasq_not_gauged}. Notice that the value of the
condensate as $\beta\to 0$ scales according to $\lambda^2$ as one might
expect from perturbation theory. The case where $\lambda=1.0$
is close to the threshold required to precipitate a condensate even in the absence
of gauge interactions and indeed we see in this case that
the condensate survives the $\beta\to\infty$ limit.

\begin{figure}[htb]\centering
    \includegraphics[width=0.45\textwidth]{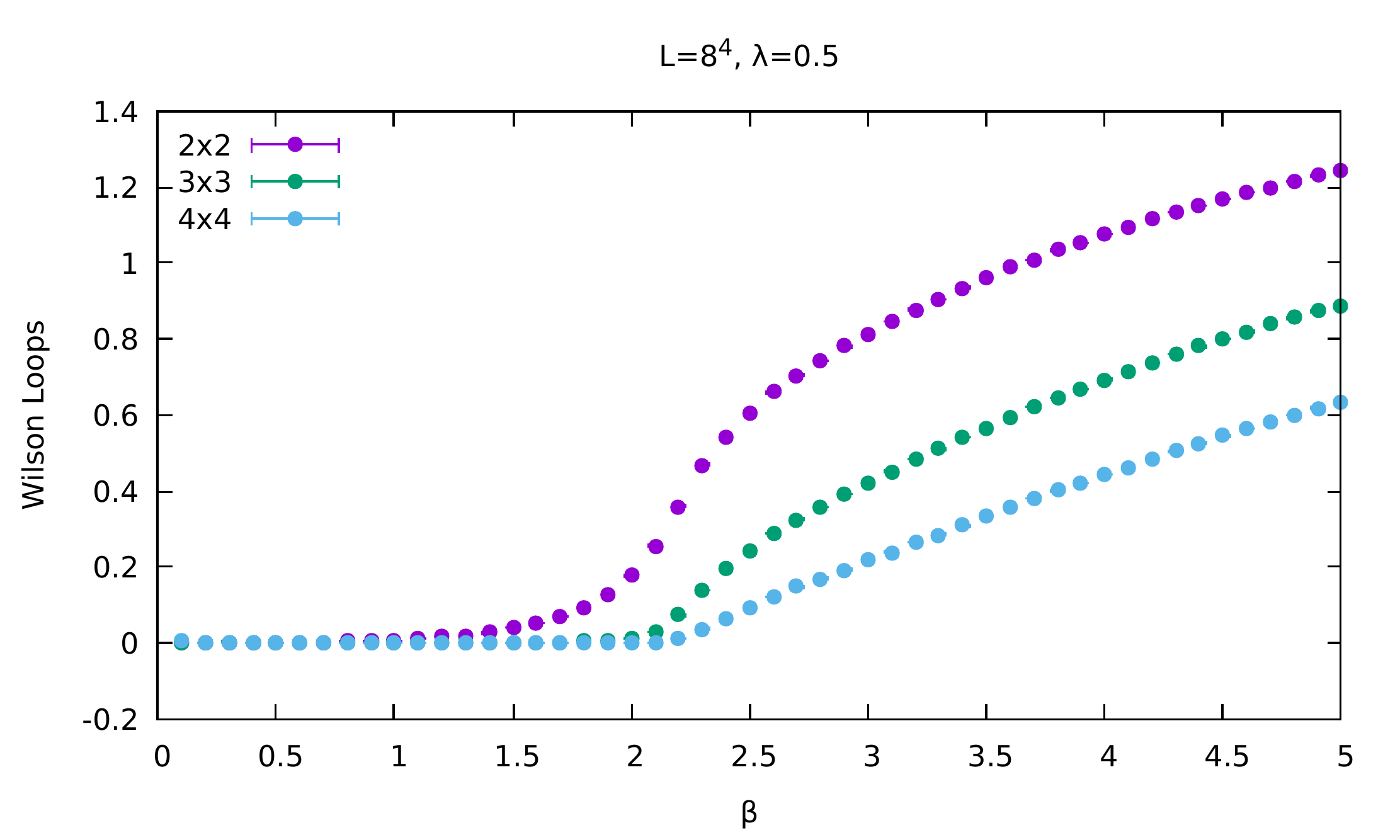}
    \caption{Wilson Loops vs $\beta$ for  $L=8^4$}
    \label{wloop_8}
\end{figure}

\begin{figure}[htb]\centering
    \includegraphics[width=0.45\textwidth]{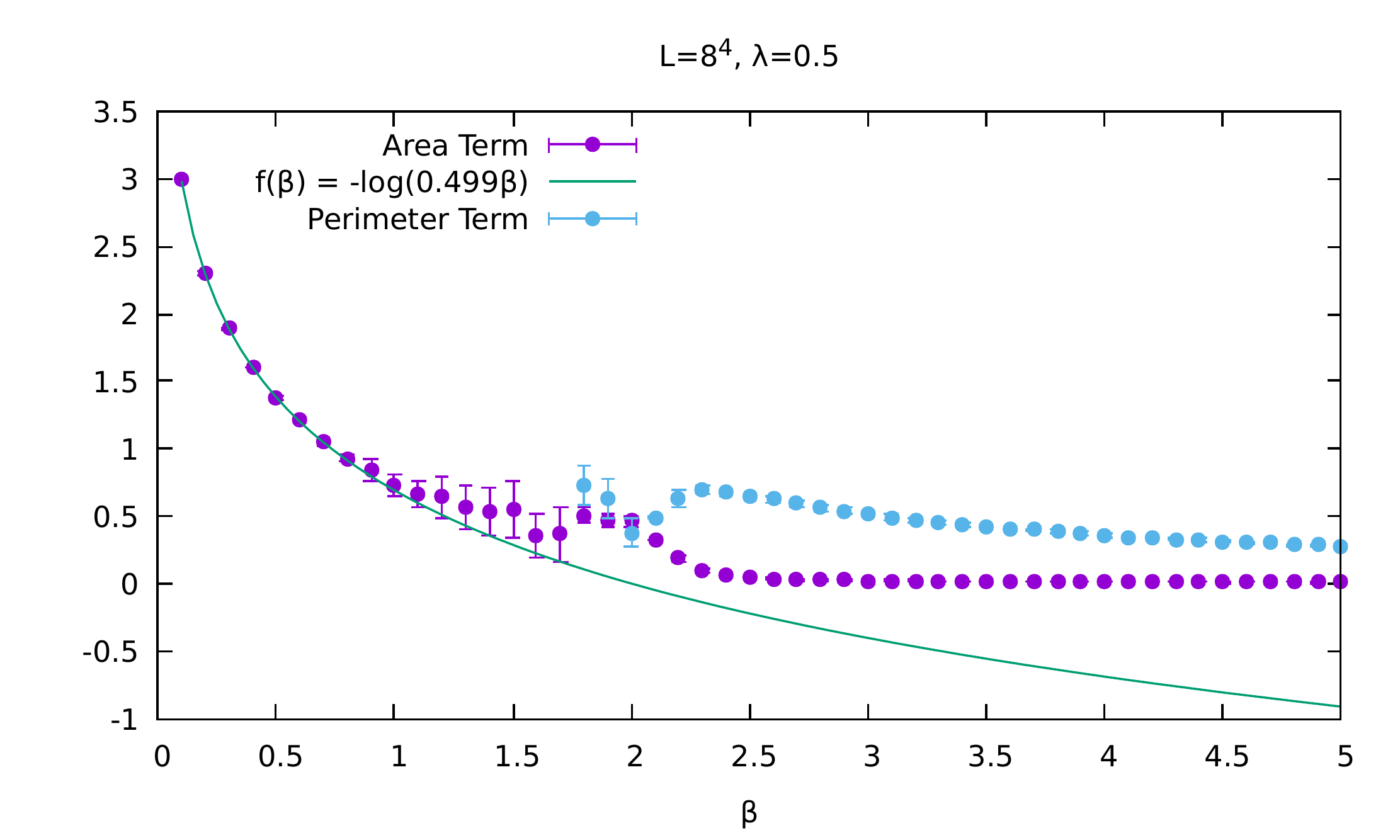}
    \caption{A and B vs $\beta$  for $L=8^4$}
    \label{st_8}
\end{figure}

The fact that the regime where the four fermion condensate is non-zero
corresponds to confinement can be seen in fig.~\ref{poly} which shows the absolute value of the Polyakov line averaged
over the lattice over the same range in $\beta$. It is clear that
the Polyakov line  vanishes for values of $\beta$ in which the four fermion
condensate grows.~\footnote{We use the absolute value of the
line in our measurements since the Polyakov line itself vanishes for all $\beta$ at finite volume as a consequence of
the exact center symmetry} A vanishing Polyakov line signals a confining phase for
the gauge theory.
This conclusion can be strengthened by looking
at Wilson loops.
The Wilson Loops for $L=8^4$ and $\lambda=0.5$ are shown in fig.\ref{wloop_8} and clearly
also decrease rapidly in the small $\beta$ regime.
To extract the string tension, we fit the $W(R,R)$ loops to an exponential of form $e^{-(AR^2+BR+C)}$ corresponding to a combination of area and perimeter laws. For values of $\beta<1.8$ the fitted values of $B$ and $C$ are consistent with zero and we hence fit only for $A$.
However around $\beta=1.8$ the area term and the perimeter term become comparable so we need to employ the full form of the exponential for couplings $\beta \ge 1.8$. This behavior can be seen in fig.\ref{st_8} which shows the coefficients $A$ and $B$ versus $\beta$.
The plot also shows the pure area law fit as a solid line
which yields an estimate of the string tension
$\sigma=0.499(5)$. This agrees well with a strong coupling analysis of the quenched
gauge theory and is consistent with the absence of light fermions in this
regime due to symmetric mass generation.

\begin{figure}[htb]\centering
    \includegraphics[width=0.45\textwidth]{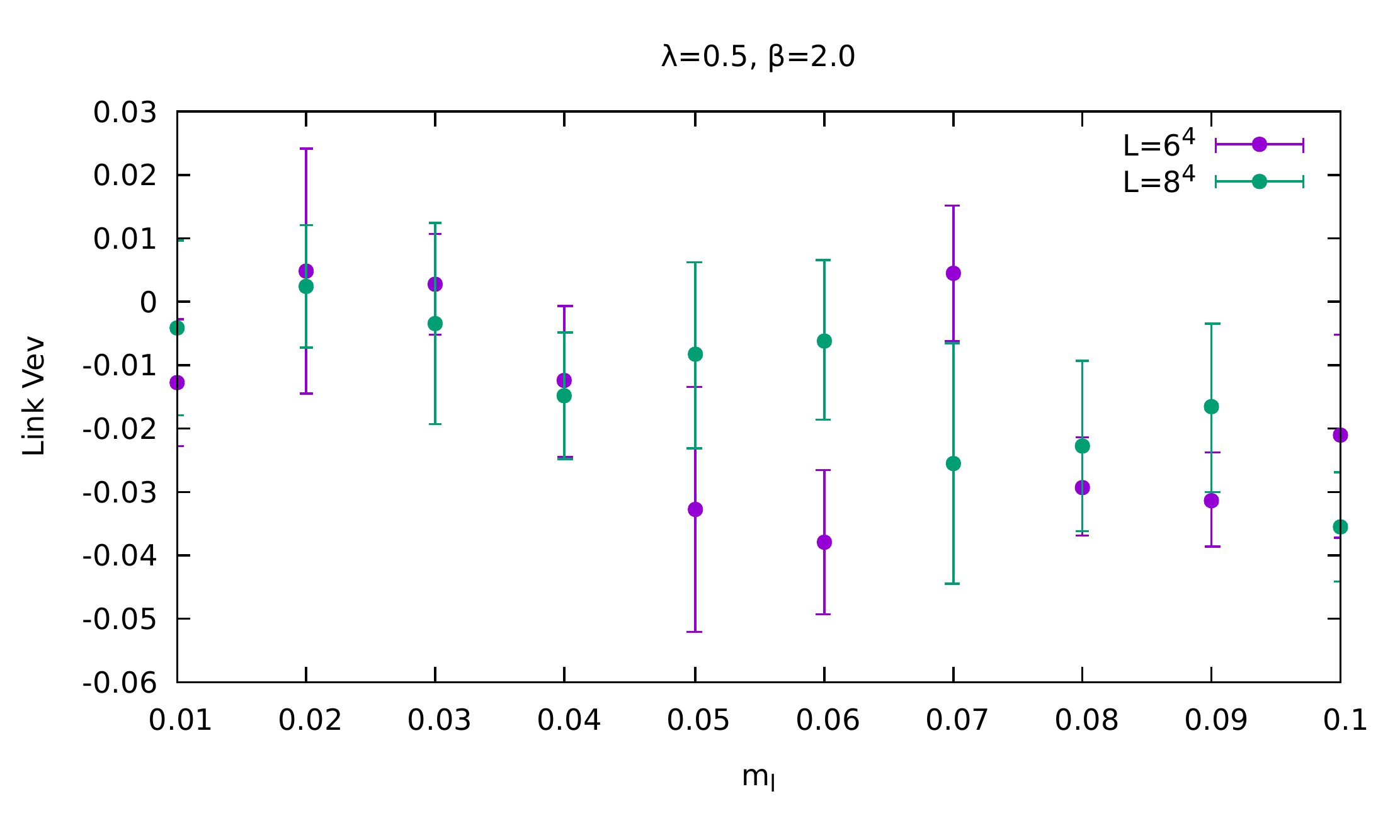}
    \caption{Link vev vs $m_l$ for $L=6^4,8^4$ at $\beta=2.0$}
    \label{link}
\end{figure}

Of course while the single site fermion bilinear is forced to vanish by symmetry in
this model
it is possible to construct other gauge invariant and $Z_4$ symmetric
fermion mass terms that involve coupling different
fermion fields within the hypercube \cite{vandenDoel:1983mf,Golterman:1984cy}.  It is
logically possible that the model would choose to condense these
other fermion bilinear operators rather than the four fermion operator
we have considered so far.
To check for this we have added the simplest of these operators, the one
link term, to the
action with coupling $m_l$.
\bea
O_1&=\frac{1}{8}\sum_{x,\mu}\epsilon(x)\xi_\mu(x){\rm Tr}[ \psi^\dagger(x)(U_\mu(x)\psi(x+\mu)
V^\dagger_\mu(x) \nonumber \\ 
&+U^\dagger_\mu(x-\mu)\psi(x-\mu)V_\mu(x-\mu)) ]
\eea
where the phase $\xi_\mu(x)=\left(-1\right)^{\sum_{\mu+1}^4 x_i}$ \cite{vandenDoel:1983mf}.
Notice though that a vev for this operator as $m_l\to 0$
will necessarily break a set of discrete shift symmetries given by
\begin{align}
    \psi(x)&\to \xi_\rho(x)\psi(x+\rho)\\
    V_\mu(x)&\to V^*_\mu(x+\rho) \\
    U_\mu(x)&\to U^*_\mu(x+\rho) 
\end{align}
In fig.~\ref{link} we show a plot of the vev of this operator for several
lattice sizes as a function of $m_l$ on a $6^4$ lattice for $\lambda=0.5$
and $\beta=2.0$. Notice that the measured vev is small in comparison with the four fermion
condensate and decreases smoothly to zero as $m_l\to 0$ with no significant dependence on lattice volume. This result argues against the condensation of such a link term and
a corresponding spontaneous breaking of these shift symmetries in the thermodynamic limit. 

\section{Conclusions}

In this paper we have argued that a particular lattice gauge theory composed of massless reduced staggered fermions transforming under a local $SU(2)\times SU(2)$ symmetry develops a four fermion rather than bilinear fermion condensate due to confinement. 
Furthermore, since this four fermion condensate breaks no symmetries there are no
Goldstone bosons in the spectrum of the theory. This gives an explicit realization of symmetric mass generation in a lattice model which describes sixteen
Majorana fermions at high energies. This number of fermion flavors
is precisely what is needed to cancel certain discrete anomalies of Weyl fermions
in the continuum \cite{Garcia-Etxebarria:2018ajm}. Our work furnishes the first
example of
of a lattice theory capable of supporting symmetric mass generation using just gauge
interactions.

\bibliography{su2}

\end{document}